\documentclass[reprint,superscriptaddress,nofootinbib,amsmath,amssymb]{revtex4-1}
\usepackage{xcolor}
\usepackage{graphicx} 
\usepackage{float}

\begin{document}
\title{General Relativistic Non-Neutral White Dwarf Stars}
	
\author{\'Erik Amorim}
\affiliation{Universit\"at zu K\"oln, Institut f\"ur Mathematik, Weyertal 86-90,
D - 50931, Cologne, Germany}
\author{Parker Hund}
\affiliation{Department of Mathematics, Rutgers University,
                110 Frelinghuysen Rd., Piscataway, NJ 08854, USA}\vspace{-10pt}
\email{\copyright (2021) The authors.\vspace{-40pt}}
\begin{abstract}
\noindent 
    We generalize the recent Newtonian two-component charged fluid models for white dwarf stars of Krivoruchenko, Nadyozhin and Yudin and of Hund and Kiessling to the context of general relativity. We compare the equations and numerical solutions of these models. We extend to the general relativistic setting the non-neutrality results and bounds on the stellar charge obtained by Hund and Kiessling.
\end{abstract}

$\phantom{xi}$\hfill 1 
\maketitle

 $\phantom{nix}$\vspace{-1.5truecm}

\section{Introduction}  \vspace{-10pt}
In \cite{KNY}, a two-component system of cold fluids in hydrostatic equilibrium with polytropic pressure density relation under Newtonian gravity was introduced and studied. This system is formed of electrons and ions and was first written
\begin{align}
    \mu_i+m_i\phi_G+Ze\phi_E &= \text{const} \ , \label{Newton1} \\
    \mu_e+m_e\phi_G-e\phi_E &= \text{const} \ , \label{Newton2}
\end{align}
where $\mu_i$ is the chemical potential of the ions, $\mu_e$ is the chemical potential of the electrons, $\phi_G$ is the gravitational potential, $\phi_E$ is the electric potential, $m_i$ and $m_e$ are the masses of the ion particles and electrons, and $Ze$ and $-e$ are their charges. To close the system, a polytropic equation of state was taken. The usual assumption \cite{Chandra} of local neutrality was not made, and this led to solutions which are not globally neutral.  

\cite{HK1} was interested in applying this model to white dwarfs, in particular using the nonrelativistic `5/3' polytropic pressure law derived from Fermi-Dirac statistics. However, they noted that, as the heavier ions in a white dwarf are bosons, such a pressure law does not apply to them. \cite{HK1} subsequently studied in greater detail the two-fluid model consisting of only electrons and protons, noting that such a model would be an approximation to a small range of `brown dwarf' stars. In that paper, it was noted that, for a wide range of pressure laws including the `5/3' and special relativistic laws discussed in \cite{Chandra}, there is a simple set of bounds on the total amount of charge such a star can hold in its ground state configuration. In CGS units, this is
\begin{equation}\label{bounds}
     \frac{1-\frac{Gm_p^2}{e^2}}{1+\frac{Gm_em_p}{e^2}}\leq\frac{N_e}{N_p}\leq\frac{1+\frac{Gm_em_p}{e^2}}{1-\frac{Gm_e^2}{e^2}} \ ,
\end{equation}
where $N_e$ and $N_p$ are the total number of electrons and protons, and $G$ is Newton's gravitational constant. See also \cite{HK2} for more details on these bounds. It was speculated that such bounds would also hold in the general relativistic case.

In the present paper, we determine the general relativistic version of these bounds, as well as present the general relativistic analogue of system (\ref{Newton1}) and (\ref{Newton2}). To do this, we apply the framework developed by Olson and Bailyn in a series of papers \cite{OB1}, \cite{OB2}, and \cite{OB3}. However, in the course of our study, we also noticed that an algebra error had been made in \cite{OB3}, the work most directly related to ours. Subsequently, the system presented there is incorrect, although it appears the correct system was used for the numerical computations. In section II we review this framework and present the corrected system of equations, in section III compare this system to the special relativistic version of (\ref{Newton1}) and (\ref{Newton2}), in section IV we extend the charge bounds results of \cite{HK1}, in section V we present the result of numerical calculations, in section VI we discuss how these results apply to neutron stars, and in section VII we conclude.

Aside from being explicitly in the vein of \cite{KNY}, \cite{HK1}, \cite{OB1}, \cite{OB2}, and \cite{OB3}, this paper can also be related to other lines of research. There is a large literature on interior solutions to the exterior Reissner-Nordström metric. Much of this work focuses on exact solutions, for example, \cite{nduka76}, \cite{nduka77}, \cite{PantSah}, \cite{PatelKoppar}, \cite{FM2013a}, \cite{FM2013b}, \cite{Kiess}, \cite{Ivanov}, \cite{Mehra1980}, \cite{HajjSfeila}, these solutions often found by ``electrifying'' a noncharged solution. \cite{REMLZ} studies the effects of charge on aspects like the mass-radius relation of stellar configurations, and even has a version of (\ref{bounds}) in their equation (24). Their model is however a single-fluid model where the charge density is proportional to the energy density, and this allows them to construct configurations with very large amounts of charge. \cite{NWMU} undertakes a similar study for strange stars, although they use a Gaussian for their charge distribution. \cite{AZ2013} and \cite{AZ2018} investigate a single fluid model with relativistic and nonrelativistic polytropic equations of state and charge density proportional to energy density,  focusing on how the charge interacts with properties such as the Buchdahl, quasiblack hole, and Oppenheimer-Volkov limits. 

\cite{Ruffini2011a} confirms the impossibility of local charge neutrality found in \cite{OB1}, as well as producing more numerical results describing the solution of the equilibrium equations. This was part of a series of papers including \cite{Ruffini2011b}, \cite{Ruffini2011c}, and \cite{Ruffini2012} which built models of white dwarfs or neutron stars using what they termed ``generalized Fermi energies'', or ``Klein potentials''. The models were taken to be globally neutral, but locally nonneutral. 

\section{The Olson-Bailyn approach}\label{OB}

In \cite{OB1}, it was noted that, in a multicomponent Einstein-Maxwell system of spherically symmetric fluids, there is one more unknown than equation. It is further noted that there are a variety of approaches to closing the system, and one of the more common is to assume local charge neutrality. Instead of making such an assumption, the authors of \cite{OB1} minimize the energy of the system to determine the last equation. In doing so, they derive an equation for every component in the fluid, which they call the ``species TOV'' equations (their sum is equivalent to the Bianchi identity for the stress-energy tensor):
\begin{equation}\label{TOV}
    \mu_i'=\frac{q_i}{r^2}\mathcal{E}e^{\lambda/2}+\mu_i\left(\frac{\lambda'}{2}-\frac{4\pi G}{c^4}re^{\lambda}\sum_j n_j\mu_j\right) \ ,
\end{equation}
where the spacetime metric is taken as
\begin{equation}\label{metric}
ds^2=e^\nu dt^2-e^\lambda dr^2-r^2(d\theta^2+\sin^2\theta d\phi^2) \ ,
\end{equation}
primes mean derivative with respect to the radial coordinate, the constant $q_i$ is the charge of each particle of the species labelled by $i$ (note that $i$ does not mean ``ion'' here, but is instead a label), $n_i(r)$ is the corresponding number density at radius $r$, $\mathcal{E}(r)$ is the total charge below radius $r$, defined via
\begin{equation} \label{calE_from_n}
    \mathcal{E}(r)=\int_0^r \left(\sum_i q_in_i(s) \right)e^{\lambda(s)/2}s^2ds \ ,
\end{equation}
and $\mu_i(r)$ is the chemical potential of the fluid of particle species $i$, defined via
\begin{equation} \label{mui_formula}
    \mu_i(r)=c^2\frac{\partial \rho_m}{\partial n_i}(r) \ ,
\end{equation}
where $\rho_m(r)$ is the total mass-energy density. It is also noted that, except under very specific circumstances, (\ref{TOV}) precludes local charge neutrality. 

In \cite{OB2}, a white dwarf star was modelled by applying these results to a two-component system formed of a perfect Fermi gas of electrons and a perfect Bose gas of nuclei. Since the Bose gas contributes no pressure and has mass-energy density of
\begin{equation}
    \rho_{m_N}=m_Nn_N
\end{equation}
($N$ for ``nucleon''), they are able to reduce the system down to a single second-order ODE, somewhat similar to the one found by Chandrasekhar in the locally neutral case. 

In \cite{OB3}, the results of \cite{OB1} were applied to a model of a neutron star consisting of three perfect Fermi gases of electrons, protons, and neutrons. The system of equations used is
\begin{widetext}
\begin{equation}\label{TOV2}
    \mu_i'=\frac{q_i\mathcal{E}}{r^2(1-2m/r)^{1/2}}+\frac{\mu_i}{1-2m/r}\left[-\frac{m}{r^2}+\frac{1}{2}\frac{G\mathcal{E}^2}{c^4r^3}+\frac{4\pi G r}{c^4}(\rho_m c^2-\sum_{j=e,p,n} n_j\mu_j)\right] \ ,
\end{equation}
\end{widetext}
where the label $i$ can be $e,p,n$ for ``electron'', ``proton'' and ``neutron'' and where $m(r)$ is defined via
\begin{equation}\label{mass}
    e^{-\lambda}=1-\frac{2m(r)}{r} \ ,
\end{equation}
together with the equations
\begin{align}
    m' &= \frac{4\pi Gr^2}{c^2}\left[\rho_m+\frac{\mathcal{E}^2}{8\pi c^2r^4}\right] \ , \label{massprime} \\
    \mathcal{E}' &= 4\pi r^2\frac{q_en_e+q_pn_p}{\left(1-\frac{2m}{r}\right)^{1/2}} \ . \label{electricprime}
\end{align}

Note that the usual mass function is related to $m$ by $M(r) = \frac{Gm(r)}{c^2}$. The mass-energy density function $\rho_m$ that appears in equation~\eqref{massprime} is assumed to be a known expression of the $n_i$, and thus also each $\mu_i$ is determined by $n_i$. In that paper and also here for us, it is assumed that $\rho_m=\rho_{m_e}+\rho_{m_p}+\rho_{m_n}$, with
\begin{multline} \label{rho_mi}
    \rho_{m_i}=\frac{\pi m_i^4c^3}{h^3}(2\overline{n}_i+\overline{n}_i^{1/3})(\overline{n}^{2/3}_i+1)^{1/2} \\
    -\ln(\overline{n}_i^{1/3}+(\overline{n}_i^{2/3}+1)^{1/2}) \ ,
\end{multline}
where 
\begin{equation}
    \overline{n}_i=\frac{3h^3n_i}{8\pi m_i^3 c^3} \ .
\end{equation}
This is the same energy density term derived in \cite{Chandra} using the special relativistic kinetic energy instead of the Newtonian kinetic energy. It yields the following relation between $\mu_i$ and $n_i$:
\begin{equation} \label{mui_SR}
    \mu_i = m_ic^2\sqrt{1+\overline{n_i}^{2/3}} \ .
\end{equation}

The strategy in \cite{OB3} to obtain a workable system for numeric computation is to first eliminate~\eqref{massprime} and~\eqref{electricprime} by solving the electron and proton equations (\ref{TOV2}) for $\mathcal{E}$ and $m$. But they have made an error here. With the definitions
\begin{equation} \label{Kdef}
    K=1-2r\frac{q_p\mu_e'-q_e\mu_p'}{q_p\mu_e-q_e\mu_p} \ ,
\end{equation}
\begin{equation}
    a=\frac{G^{1/2}r}{c^2}\frac{\mu_e\mu'_p-\mu'_e\mu_p}{q_p\mu_e-q_e\mu_p} \ ,
\end{equation}
\begin{equation}
    b=\frac{8\pi G r^2}{c^4}(\rho_m c^2-\sum_i n_i\mu_i)
\end{equation}
(equations (3.7)--(3.9) in \cite{OB3}), they write that
\begin{equation}
    m=\frac{r}{2}\frac{b+a^2}{K+a^2} \ ,
\end{equation}
\begin{equation}
    \mathcal{E}=\frac{c^2ra}{G^{1/2}}\left(\frac{K-b}{K+a^2}\right)^{1/2}
\end{equation}
(equations (3.5) and (3.6) in \cite{OB3}), while in fact the correct equations are
\begin{equation}
    m=\frac{r}{2}\frac{b+a^2+K-1}{K+a^2} \ ,
\end{equation}
\begin{equation}
    \mathcal{E}=\frac{c^2ra}{G^{1/2}}\left(\frac{1-b}{K+a^2}\right)^{1/2} \ .
\end{equation}
Consequently, from (\ref{massprime}) and (\ref{electricprime}) they derive
\begin{align}
    \frac{d}{dr}\left(r\frac{b+a^2}{K+a^2}\right) &= \frac{8\pi Gr^2}{c^2}\rho_m+a^2\frac{K-b}{K+a^2} \ , \label{OB3equation1} \\
    \frac{d}{dr}\left(r^2a^2\frac{K-b}{K+a^2}\right) &= \frac{8\pi G^{1/2}ar^3}{c^2}(q_en_e+q_pn_p) \label{OB3equation2}
\end{align}
(equations (3.10) and (3.11) in \cite{OB3}), while the correct equations should be
\begin{align}
    \frac{d}{dr}\left(r\frac{b+a^2+K-1}{K+a^2}\right) &= \frac{8\pi G r^2}{c^2}\rho_m+a^2\frac{1-b}{K+a^2} \ , \label{actualMassPrime} \\
    \frac{d}{dr}\left(r^2a^2\frac{1-b}{K+a^2}\right) &= \frac{8\pi G^{1/2}ar^3}{c^2}(q_en_e+q_pn_p) \ . \label{actualElectricPrime}
\end{align}
Note that the corrected equations would not differ from the ones they reported if $K=1$, while in fact one can see from~\eqref{Kdef} that, for small $r$, the value of $K$ is close to but not equal to 1.

Next, it is explained in \cite{OB3} how to produce numerical solutions to this model by first noting that $n_p-n_e$ will be very small, about $10^{-33}$, so that it can be assumed that $n_p=n_e$ and (\ref{OB3equation1}) can be solved, taking for initial conditions at $r=0$ an arbitrary value of the central density $n_p(0)=n_e(0)$ as well as $n_p'(0)=n_e'(0)=0$ (which is dictated by spherical symmetry). The neutron variables $n_n$ and $\mu_n$, which appear in (\ref{OB3equation1}), are determined by the $\beta$-equilibrium condition $\mu_n=\mu_p+\mu_e$ (equation (2.12) in \cite{OB3}). The result is then plugged into (\ref{OB3equation2}) to get an estimate of the total charge.

There are two remarks that we would like to make about this procedure. The first is that it appears that the numerical calculations in \cite{OB3} were done using the correct equations. Indeed, the apparent singularity at $r=0$ is removable for the corrected system (\ref{actualMassPrime}) and (\ref{actualElectricPrime}), but not for the originally reported equation (\ref{OB3equation1}) (see the appendix for this calculation), so it would not be possible to choose the central density as initial condition for the equations as they appear in that paper. But we have redone the computations using the corrected equations and found identical plots to theirs.

The second remark is that the system comprised of equations~\eqref{actualMassPrime} and~\eqref{actualElectricPrime} is not useful when one wants to consider $n_p\neq n_e$, as we want to do. This is because this system was derived using (\ref{TOV2}) for both values $i=e,p$, which is only valid when $n_i\neq0$. Indeed, it is implicit in the model that, for each label $i$, if $n_i(r_0)=0$, then we set $n_i(r)=0$ for all $r\geq r_0$. In particular, in the case of only two particle species, if $r_0$ is the radius at which one of the two particle densities (say, the electrons) first reaches zero, we solve for the protons by numerically solving the system given by the protons' (\ref{TOV2}), (\ref{massprime}), and (\ref{electricprime}), choosing the boundary conditions so as to have continuity, but (\ref{TOV2}) for the electrons is not valid for $r\geq r_0$. Therefore, outside of the support of one of the particle species,~\eqref{actualMassPrime} and~\eqref{actualElectricPrime} are not valid.

\section{Comparisons with the nonrelativistic model}

In this section we only consider models with electrons and protons, with the goal of drawing a comparison between the general relativistic model described in section \ref{OB} (``the GR model'') and the special relativistic one described by (\ref{Newton1}) and (\ref{Newton2}) using the special relativistic kinetic energy to derive relation~\eqref{mui_SR} between the chemical potential and the particle density (``the SR model''). The latter model was given in \cite{HK1} in the form
\begin{widetext}
\begin{align}
    k_py''_p = -k_p\frac{2}{r}y_p'+\frac{1}{l_p^3}\left(1-\frac{Gm_p^2}{q_p^2}\right)(y_p^2-1)^{3/2}-\frac{1}{l_e^3}\left(1+\frac{Gm_pm_e}{q_p^2}\right)(y^2_e-1)^{3/2} \ , \label{special1} \\
    k_ey''_e=-k_e\frac{2}{r}y_e'+\frac{1}{l_e^3}\left(1-\frac{Gm_e^2}{q_p^2}\right)(y_e^2-1)^{3/2}-\frac{1}{l_p^3}\left(1+\frac{Gm_pm_e}{q_p^2}\right)(y^2_p-1)^{3/2} \ . \label{special2}
\end{align}
\end{widetext}
where $l_i=\frac{(3\pi^2)^{1/3}\hbar}{m_ic}$, $k_i=\frac{m_ic^2}{4\pi q_p^2}$, and $y_i=\sqrt{1+l_i^2n_i^{2/3}}$. So, in fact, the unknowns $y_e$ and $y_p$ are related to the chemical potentials by $\mu_i=m_ic^2y_i$. Rewriting this system in terms of $n_i$ and $\mu_i$, we obtain
\begin{align}
    (r^2\mu_p')' &= +4\pi q_p^2 r^2 (n_p-n_e) - 4\pi G m_p r^2 (m_pn_p+m_en_e) \ , \label{new_special1} \\
    (r^2\mu_e')' &= -4\pi q_p^2 r^2 (n_p-n_e) - 4\pi G m_e r^2 (m_pn_p+m_en_e) \ . \label{new_special2}
\end{align}

Meanwhile, the main equations of the GR model are the two equations~\eqref{TOV2} for $\mu_p$ and $\mu_e$ (taking $\mu_n=n_n=0$ on the right hand side), equation~\eqref{massprime} for the mass function $m$, and equation~\eqref{electricprime} for the electric charge $\mathcal{E}$, together with the definitions~\eqref{rho_mi} and~\eqref{mui_formula} of the energy density $\rho_m = \rho_{m_e}+\rho_{m_p}$ and the chemical potentials $\mu_i$. Instead of considering $m$, we will work here with the true mass function $M(r) = c^2m(r)/G$. We also remark that~\eqref{massprime} can be used to rewrite~\eqref{TOV2} in the following form:
\begin{widetext}
\begin{equation} \label{new_TOV2}
    r^2\mu_i'=\frac{q_i\mathcal{E}}{(1-\frac{2GM}{c^2r})^{1/2}}+\frac{\mu_i}{1-\frac{2GM}{c^2r}}\left[r^2\left(\frac{GM}{c^2r}\right)'-\frac{4\pi G r^3}{c^4}(n_e\mu_e+n_p\mu_p) \right] \ .
\end{equation}
\end{widetext}

\subsection{The post-Minkowski approximation}

A natural question to ask is whether the SR and GR models coincide in low orders of a power series expansion in $G$. We shall see here that they match only in zeroth order, which is to be expected: on the one hand, when $G=0$, the flat space of special relativity also solves the Einstein equations of general relativity, thus explaining the match at zeroth order; on the other hand, the SR model contains terms of degree 1 in $G$ that are nonrelativistic in nature, derived from the Newtonian potential, and there's no reason to hope that they will result from any approximation scheme performed from a fully relativistic model.

For $G=0$, equations~\eqref{new_special1} and~\eqref{new_special2} of the SR model reduce to
\begin{align}
    (r^2\mu_p')' &= +4\pi q_p^2 r^2 (n_p-n_e) \ , \label{postmink_1} \\
    (r^2\mu_e')' &= -4\pi q_p^2 r^2 (n_p-n_e) \label{postmink_2} \ ,
\end{align}
while equation~\eqref{new_TOV2} of the GR model becomes the following two when written for $i=p$ and $i=e$:
\begin{align}
    r^2\mu_p' &= q_p\mathcal{E} \label{postmink_3} \ , \\
    r^2\mu_e' &= q_e\mathcal{E} \ . \label{postmink_35}
\end{align}
Considering also equation~\eqref{electricprime} for $\mathcal{E}$ in the GR model, which when $G=0$ reduces to $\mathcal{E}' = 4\pi r^2 (q_en_e+q_pn_p)$, we see that equations~\eqref{postmink_1} and~\eqref{postmink_2} yield the same solutions as~\eqref{postmink_3} and~\eqref{postmink_35} provided the chosen initial conditions $n_i(0)$ and $n_i'(0)=0$ are the same between the models.

To study what happens for $G\neq 0$, we let $n_i^0$ and $\mu_i^0$ denote the solutions to the $G=0$ equations of either model, and we also use a superscript zero on any expression that depends on these solutions, like $\rho_m^0$, $\mathcal{E}^0$ etc. We also write the solutions of either the SR or the GR model for $G\neq 0$ in the form
\begin{equation}
    \mu_i = \mu_i^0 + G\mu_i^1 + O(G^2) \ , \quad n_i = n_i^0 + Gn_i^1 + O(G^2) \ ,
\end{equation}
and apply the same notation to any expression that depends on $n_i$ and $\mu_i$. Working only to first order in $G$, it is immediate that the SR model equations imply
\begin{align}
    \hspace{-1em} (r^2(\mu_p^1)')' &= +4\pi q_p^2r^2(n_p^1-n_e^1) - 4\pi m_p r^2(m_pn_p^0+m_en_e^0) , \label{postmink_5} \\
    \hspace{-1em} (r^2(\mu_e^1)')' &= -4\pi q_p^2r^2(n_p^1-n_e^1) - 4\pi m_e r^2(m_pn_p^0+m_en_e^0) . \label{postmink_6}
\end{align}
Meanwhile, keeping only first order terms in equation~\eqref{new_TOV2} of the GR model gives
\begin{multline} \label{postmink_4}
    r^2(\mu_i^1)' = q_i\mathcal{E}^1 + \frac{q_i\mathcal{E}^0M^0}{c^2r} \\
    + \mu_i^0\left[r^2\left(\frac{M^0}{c^2r}\right)'-\frac{4\pi r^3}{c^4}(n_e^0\mu_e^0+n_p^0\mu_p^0)\right] \ ,
\end{multline}
where $\mathcal{E}^1$ and $M^0$ are obtained from the first order version of equations~\eqref{electricprime} and~\eqref{massprime}:
\begin{align}
    \hspace{-1em} (\mathcal{E}^1)' &= 4\pi r^2 \left( \frac{M^0}{c^2r}(q_en_e^0+q_pn_p^0) + q_en_e^1+q_pn_p^1 \right) \ , \\
    \hspace{-1em}  (M^0)' &= 4\pi r^2 \rho_m^0 + \frac{\mathcal{E}^0\mathcal{E}^1}{c^2r^2} \ .
\end{align}
After differentiating~\eqref{postmink_4}, we don't find agreement with the SR equations~\eqref{postmink_5} and~\eqref{postmink_6}, nor should we expect to, considering that the mass density $\rho_m$, which enters the equations of the GR model but not those of the SR model, can at this point still be any arbitrary function of the $n_i$.

\subsection{The weak field limit of the GR model}

A different type of approximation scheme often performed in general relativity is the \emph{weak field limit}, in which the metric is assumed to be close to the flat metric of special relativity. One might ask whether the GR model reduces to the SR model under this approximation, but we shall see here that the answer is also no, although the discrepancy is small.

Taking into account the definition~\eqref{mass} of the mass function, we see that, in order to find the weak field limit of the GR model, one must assume that the expression $GM/c^2r$ and its derivatives are small. So let us study what the equations of the model become when neglecting all occurrences of $GM/c^2r$ and its derivative in~\eqref{electricprime} and~\eqref{new_TOV2}. This amounts to studying the zeroth order terms in our equations with respect to an expansion in powers of $GM/c^2r$. The relevant equations become:
\begin{align}
    \mathcal{E}' &= 4\pi r^2(q_en_e+q_pn_p) \ , \label{E_wfl} \\
    r^2\mu_i' &= q_i\mathcal{E} -\frac{4\pi G \mu_i r^3}{c^4}(n_e\mu_e+n_p\mu_p) \ . \label{mui_wfl}
\end{align}
Therefore, taking the derivative of~\eqref{mui_wfl} and using~\eqref{E_wfl} on the right hand side,
\begin{multline}
    (r^2\mu_i')' = 4\pi q_i r^2 (q_en_e+q_pn_p) \\
    -\frac{4\pi G}{c^4}\bigg(\mu_i r^3(n_e\mu_e+n_p\mu_p)\bigg)' \ .
\end{multline}
This only matches the SR equations~\eqref{new_special1} and~\eqref{new_special2} in the contribution from the electric charge, that is, the first term on the right hand side.

\subsection{The weak field nonrelativistic limit}

Let us now check that, assuming a weak field and under the Newtonian limit $c\to\infty$, the GR and SR models reduce to the Newtonian model described in \cite{KNY}. This model is comprised of~\eqref{Newton1} and~\eqref{Newton2}, but using the nonrelativistic form of the kinetic energy to derive the following relation between the chemical potential and the particle density:
\begin{equation}
    \mu_i^{NR} = \frac{h^2}{2m_i}\left(\frac{3n_i}{8\pi}\right)^{2/3} \ .
\end{equation}
This corresponds to the $c$-independent term in the expansion with respect to powers of $c^2$ of the relativistic $\mu_i$ given in~\eqref{mui_SR}:
\begin{equation}
    \mu_i = m_ic^2 + \frac{h^2}{2m_i}\left(\frac{3n_i}{8\pi}\right)^{2/3} + O\left(\frac{1}{c^2}\right) \ .
\end{equation}
If we disregard the terms containing powers of $c^{-2}$ from this and plug it into the $\mu_i'$ terms of the SR model equations~\eqref{new_special1} and~\eqref{new_special2}, the constants $m_ic^2$ will drop out and we easily find the Newtonian model.

As for the GR model, we begin with the equation
\begin{equation} \label{generalized_mu}
    e^{\nu/2}\mu_i + q_i\phi_E = \text{const} \ ,
\end{equation}
where
\begin{equation}
    \phi_E(r) = \int_0^r \frac{\mathcal{E}(s)}{s^2}e^{(\lambda(s)+\nu(s))/2} ds
\end{equation}
is the electric potential. Equation~\eqref{generalized_mu} was given in section II of \cite{OB3} and is equivalent to the STOV equations~\eqref{TOV2}. Now define as in \cite{Ruffini2011c} the Newtonian gravitational potential
\begin{equation}
    \phi_G(r) = \frac{c^2\nu(r)}{2} \ .
\end{equation}
Note that
\begin{equation}
    e^\nu \approx 1 + \frac{2\phi_G}{c^2} \ .
\end{equation}
Hence, in the weak field limit, we may disregard high order terms in $\phi_G$. We can find a differential equation for $\phi_G$ by considering the radial component of the Einstein equations, which reads
\begin{equation}
    \phi_G' = G\frac{M-\frac{4\pi r^3P}{c^2}-\frac{\mathcal{E}^2}{2rc^2}}{r^2\left(1-\frac{2GM}{c^2r}\right)} \ ,
\end{equation}
where $P$ is the pressure. If we keep in all equations of the model only terms of up to first order in $\phi_G$ as well as $GM/c^2r$, as well as let $1/c^2=0$ in order to obtain the Newtonian limit, then the equation above turns into the usual Newtonian equation
\begin{equation}
    \phi_G' = \frac{GM}{r^2} \ ,
\end{equation}
where $M$ now satisfies
\begin{equation}
    M'(r)=4\pi r^2 (m_en_e+m_pn_p) \ .
\end{equation}
We also find that the electric potential reduces to
\begin{equation}
    \phi_E' = \frac{\mathcal{E}}{r^2} \ ,
\end{equation}
which is the same as its usual classical equation, considering the weak field form~\eqref{E_wfl} of the $\mathcal{E}$ equation. Now replace $e^{\nu/2}$ by $1+\phi_G/c^2$ and $\mu_i$ by $m_ic^2+\mu_i^{NR}$ in~\eqref{generalized_mu}, multiply out, ignore the ensuing term $\mu_i^{NR}/c^2$ and differentiate; we find the Newtonian model equations
\begin{equation}
    (\mu_i^{NR})' + m_i\phi_G' + q_i\phi_E' = 0 \ .
\end{equation}

\section{Bounds on the charge}\label{boundsSection}

In this section we find charge bounds analogous to (\ref{bounds}) for the GR model of a system composed of electrons and protons. Assume that we have a solution of the system and suppose that the support of $n_p$ is larger than that of $n_e$ and is bounded (if it has unbounded support, much of the next section can be skipped, although we do need to assume sufficient decay of $\mu_p'$). Then multiply the proton version of (\ref{TOV2}) by $r^2$ and evaluate at $R$, the radius of the star, which by definition is the smallest $r$ at which $n_e(r)=n_p(r) = 0$. We get
\begin{widetext}
\begin{equation}\label{GRbound1}
    \frac{q_p\mathcal{E}(R)}{\left(1-\frac{2m(R)}{R}\right)^{1/2}}-\frac{m_pc^2m(R)}{1-\frac{2m(R)}{R}}+\frac{m_pc^2}{1-\frac{2m(R)}{R}}\frac{G\mathcal{E}^2(R)}{2c^4R}= R^2\mu'_p(R) \ .
\end{equation} 
\end{widetext}
Let $Q=\mathcal{E}(R)$ be the total charge and $M=\frac{c^2}{G}m(R)$ the total mass. Then (\ref{GRbound1}) can be written as
\begin{equation}\label{GRbound2}
    \frac{q_pQ}{\left(1-\frac{2GM}{c^2R}\right)^{1/2}}-\frac{m_pGM}{1-\frac{2GM}{c^2R}}+\frac{m_pc^2}{1-\frac{2GM}{c^2R}}\frac{GQ^2}{2c^4R}= R^2\mu'_p(R) \ .
\end{equation}
If it was the electron support which was larger, the same argument would give us
\begin{equation}\label{GRbound3}
    \frac{q_eQ}{\left(1-\frac{2GM}{c^2R}\right)^{1/2}}-\frac{m_eGM}{1-\frac{2GM}{c^2R}}+\frac{m_ec^2}{1-\frac{2GM}{c^2R}}\frac{GQ^2}{2c^4R}=R^2\mu'_e(R)
\end{equation}
(notice that now the first term is negative). These equations are the analogues of equations (67) and (70) from \cite{HK1}. 

Since (\ref{bounds}) gives bounds on the \emph{most extreme} charges a configuration can hold, to make a direct comparison we do the same. However, since there are two degrees of freedom in the GR model (in both cases the central densities), what we call ``the most extreme charge'' is only most extreme under the assumption of some other quantity being fixed. In the nonrelativistic case, this quantity was $N_p = N_p(R)$, which was an easy quantity to work with since it was directly related to both $M$ and $Q$ via $M=m_eN_e+m_pN_p$ and $Q=q_eN_e+q_pN_p$. It was noted in \cite{HK1} that the density $n_p$ or $n_e$ must be decreasing at the boundary of the star so that the right hand sides of (67) and (70) in that paper are both nonpositive, which gives
\begin{align}
    q_eQ-m_eGM &\leq 0 \ , \label{bounds2} \\
    q_pQ-m_pGM &\leq 0 \ , \label{bounds3}
\end{align}
and plugging the definitions of $M$ and $Q$ in terms of $N_e$ and $N_p$ into these equations yields the charge bounds (\ref{bounds}). In the GR model, however, $M$ is not defined as a simple linear combination of $N_e$ and $N_p$, so $N_p$ is not easy to work with. We therefore seek to make a comparison to (\ref{bounds2}) and (\ref{bounds3}), which we can see already somewhat resemble (\ref{GRbound2}) and (\ref{GRbound3}).

To this end, we first also make the observation as in \cite{HK1} that $\mu_p'(R)\leq 0$, since $\mu_p$ decreases to $m_pc^2$ close to the surface of the star. Now assume that we have a positively charged configuration, that is, $Q>0$. We are interested in the most charged configurations, so we consider what happens as we increase $N_p$ and hold $N_e$ fixed. If we knew that the radius $R$ diverges faster than $Q$ or $M$ as $N_p$ increases, we could apply this to (\ref{GRbound2}) to give us (\ref{bounds3}). We therefore seek to prove this.

As explained in section III of \cite{OB1}, the special TOV equations are obtained from the condition $\frac{\partial\Lambda}{\partial N_i}=0$ for
\begin{equation}
    \Lambda=M+\frac{GQ^2}{Rc^4} \ .
\end{equation}
Differentiating this with respect to $N_i$ and taking into account $Q = q_eN_e+q_pN_p$, we can compute that 
\begin{equation}\label{rgrowth}
    \frac{\partial R}{\partial N_i}=\frac{2Rq_i}{Q}+\frac{c^4R^2}{GQ^2}\frac{\partial M}{\partial N_i} \ .
\end{equation}
Since we are assuming $Q>0$ and are increasing $N_p$, the first term on the right is positive. So, if we can show that $\frac{\partial M}{\partial N_p}>0$, we show that the radius is increasing with increasing $N_p$. We show in the appendix that


\begin{equation} \label{partialM_partialNp}
    \frac{\partial M}{\partial N_p} = m_pe^{-\lambda(R)/2} \ .
\end{equation}
Therefore we can conclude that, given $Q>0$, as $N_p$ increases, so too does $R$.

Now we would like to conclude that $R$ increases faster than $Q$ and $M$. First note that, by (\ref{rgrowth}), $R$ grows faster than $M$ since $\frac{c^4R^2}{GQ^2}>1$. By (\ref{mass}) we have 
\begin{equation}
    e^{-\lambda(R)}=1-\frac{2M}{R},
\end{equation}
so as $R$ increases, $e^{-\lambda(R)/2}$ increases, this allows us to say that as we increase $R$, we have a lower bound on $e^{-\lambda(R)/2}$, let us call it $e^{-\overline{\lambda}/2}$. So we can say that in this situation, 
\begin{equation}
    \frac{\partial M}{\partial N_p} \geq m_pe^{-\overline{\lambda}/2}>0 \ .
\end{equation}

Since $Q$ grows linearly in $N_p$, we can write that $Q(N_p)=Q_0+q_pN_p$ for an appropriate $Q_0$. Define $\hat{R}$ through
\begin{equation}\label{rhat}
    \frac{\partial \hat{R}}{\partial N_p}=\frac{c^4\hat{R}^2m_p}{G(Q_0+q_pN_p)^2}e^{-\overline{\lambda}/2} \ .
\end{equation}
Comparing (\ref{rgrowth}) and (\ref{rhat}), we can see that $R>\hat{R}$, as long as we choose the correct initial condition for $\hat{R}$. (\ref{rhat}) can be solved by a separation of variables to give 
\begin{equation}
    \frac{1}{\hat{R}}=\frac{c^4m_pe^{-\overline{\lambda}}}{Gq_p(Q_0+N_pq_p)}+K \ .
\end{equation}
Now $K$ is negative if
\begin{equation}
    \hat{R}(0)>\frac{Gq_pQ_0}{c^4m_pe^{-\overline{\lambda}/2}} \ .
\end{equation}
Since we can ``start'' the differential equation when $Q_0=q_p>0$, this inequality will be true as the right hand side is very small, but the left is larger than the radius of a neutral configuration, which is still thousands of kilometers. Then 
\begin{equation}
    \hat{R}=\frac{Gq_p(Q_0+N_pq_p)}{c^4m_pe^{-\overline{\lambda}/2}+KGq_p(Q_0+N_pq_p)} \ .
\end{equation}
Since $K$ is negative, $\hat{R}$ blows up at a finite $N_p$, which implies that so does $R$. We may then conclude that, for fixed $N_e$, $Q$ is bounded above and $R$ grows faster than $Q$ for large enough $N_p$.

We can get another lower bound on $R$ by following the same procedure with the first term on the right in (\ref{rgrowth}). Define $\bar{R}$ by
\begin{equation}
     \frac{\partial \bar{R}}{\partial N_p}=\frac{2\bar{R}q_p}{Q_0+N_pq_p} \ .
\end{equation}
Solving this we get 
\begin{equation}
    \bar{R}=(Q_0+q_pN_p)^2+K
\end{equation}
for $K$ the integration constant. We can therefore conclude that $R>Q$.

 Thus we can finally conclude that, as $N_p$ grows, $R$ grows faster than $M$ and $Q$, and further that there is some finite $N_p^\infty$ such that $R\rightarrow \infty$ as $N_p\rightarrow N_p^\infty$. Therefore, by taking this limit, we can determine that, in the most positively charged cases, (\ref{GRbound2}) reduces to (\ref{bounds3}). The exact same argument works to show that, in the most negatively charged cases, (\ref{GRbound3}) reduces to (\ref{bounds2}). So we conclude that, in terms of $Q$ and $M$, the relativistic bounds on charge are the same as~\eqref{bounds2} and~\eqref{bounds3}.

\section{Numerical Results}
We can directly compare the solutions of our system to the numerical results found in \cite{HK1}, reproduced here for ease. For reasons which are explained there, it is very difficult to produce meaningful graphs using the true physical values of the constants. We therefore adopt the same approach of using `science fiction' values $m_e/m_p=1/10$ and $Gm_p^2/e^2=1/2$ (where $e=q_p=-q_e$).

Figures \ref{53} and \ref{SR} are sample plots using the Newtonian set of equations (\ref{Newton1}) and (\ref{Newton2}) and the special relativistic set of equations (\ref{special1}) and (\ref{special2}), respectively. These, and the ones like them found in \cite{HK1}, can be compared to the plots in figures \ref{GRneg} and \ref{GRpos}. In all cases, the plots were made by holding the central density of the protons constant while slightly altering the central density of the electrons. One can see that the relativistic solutions have the same general structure as the nonrelativistic solutions: as the electron central density is increased, the electron radius approaches infinity, while, as the electron central density is decreased, the proton radius approaches infinity.

\begin{figure}[ht]
 \includegraphics[scale=0.7]{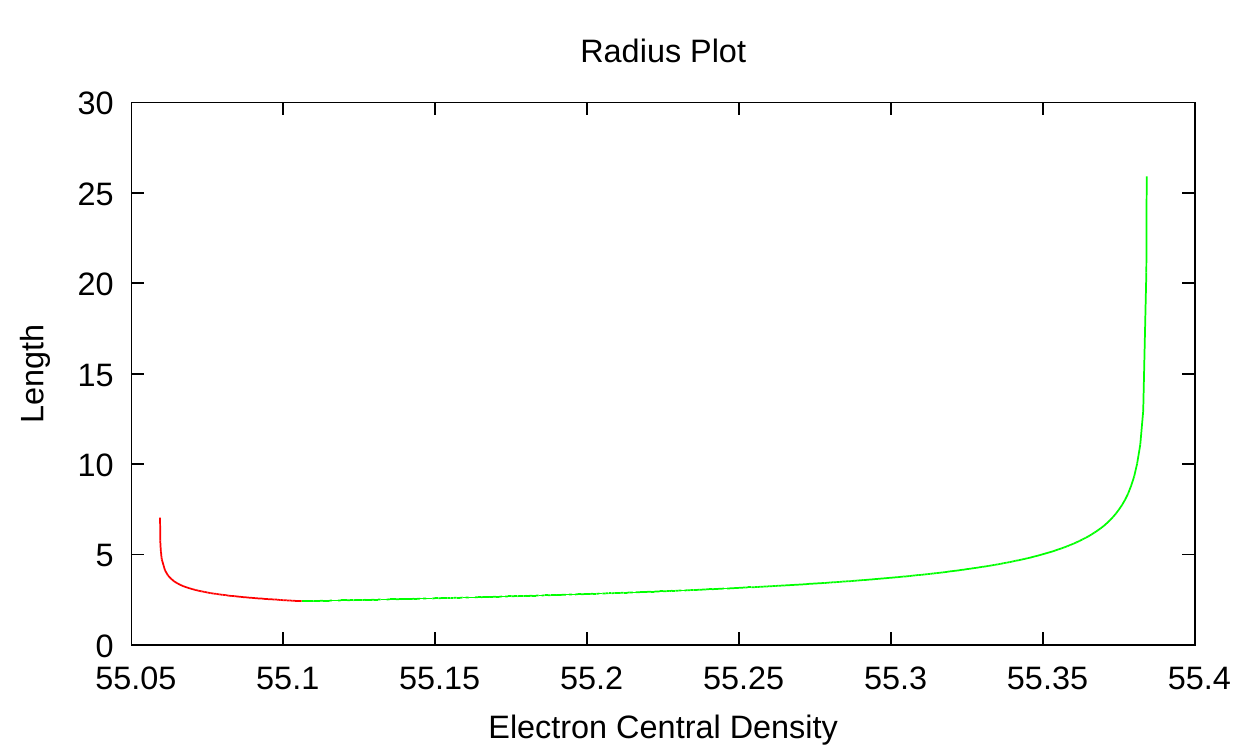}
\caption{Shown are the radii of the configurations as the central electron density is varied while the central proton density is held fixed at 100. The red colored part of the curve represents configurations with a positively charged atmosphere while the green part of the curve represents configurations with a negatively charged atmosphere. The science fiction values of $Gm_p^2/e^2=1/2$ and $m_e/m_p=1/10$ are used, so the values of length are not meaningful.}
\label{radiusplot}
\end{figure}

\begin{figure}[H]
 \includegraphics[scale=0.7]{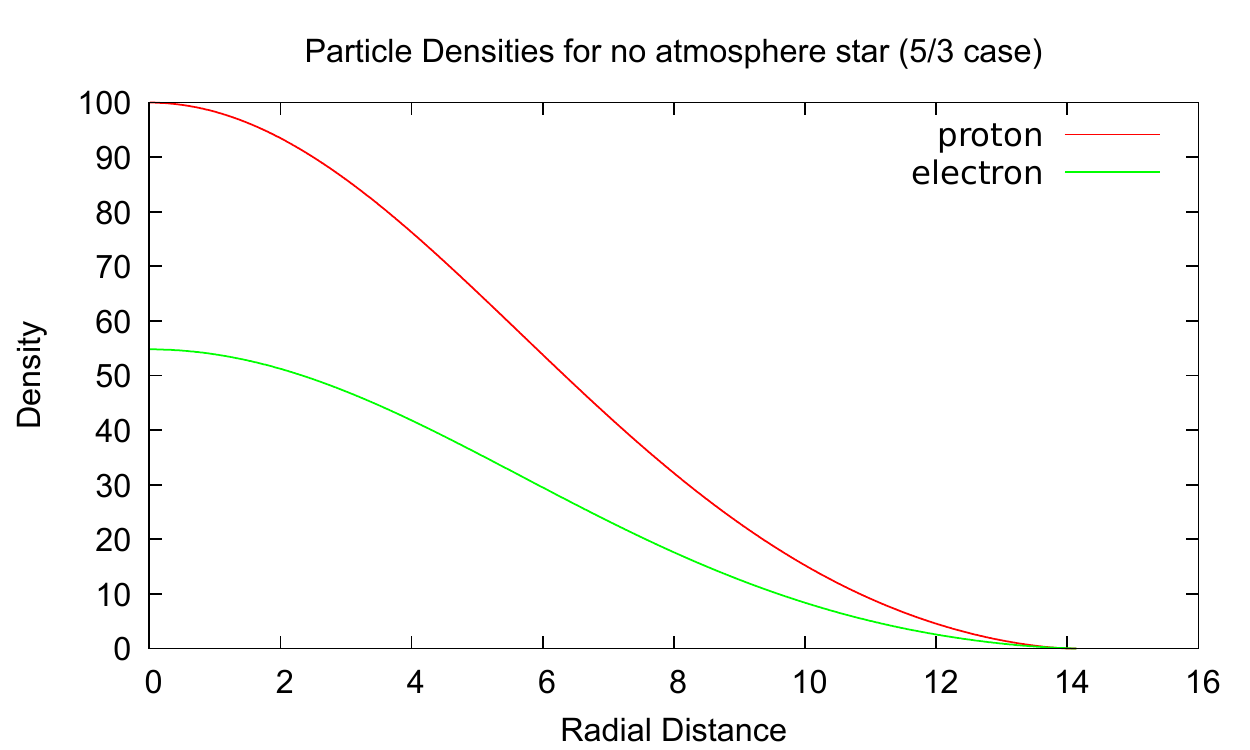}
\caption{An example when the kinetic energy law is 5/3.}
\label{53}
\end{figure}

\begin{figure}[H]
 \includegraphics[scale=0.7]{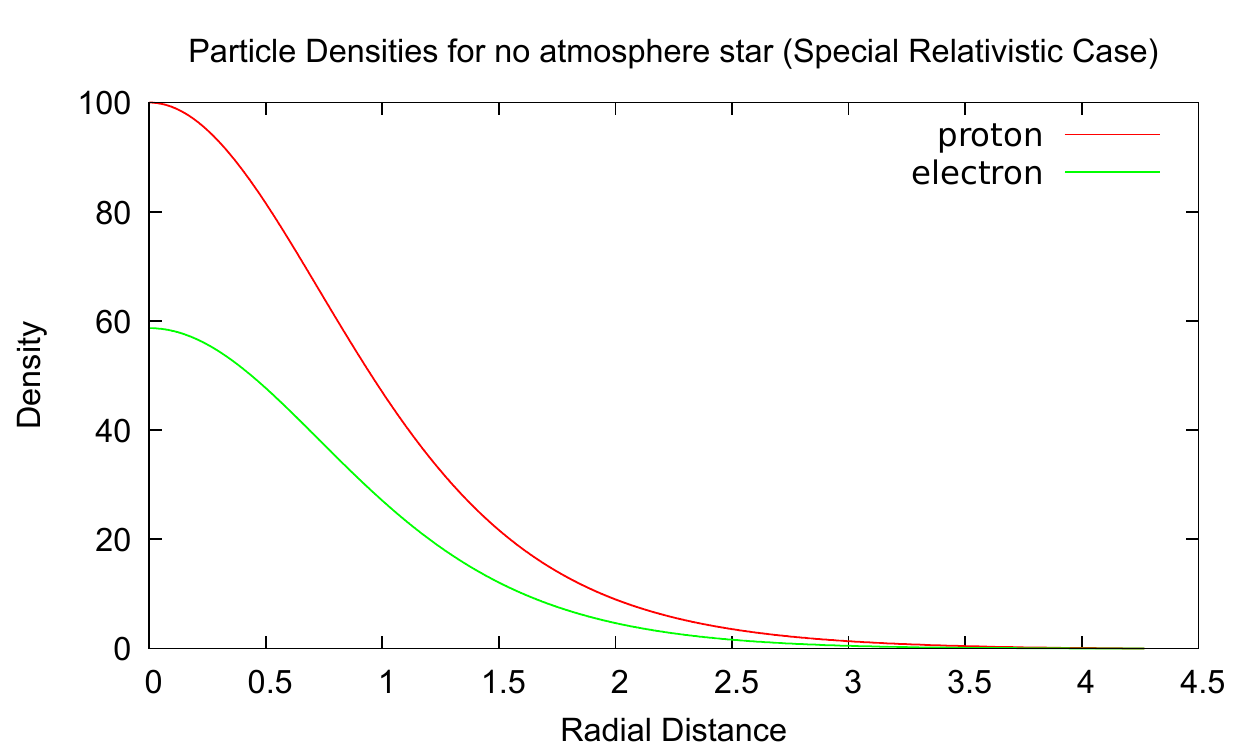}
\caption{An example when the kinetic energy law is special relativistic.}
\label{SR}
\end{figure}

 The radius is of course difficult to see since the densities themselves quickly decay to be very close to zero. Figure \ref{radiusplot} is a plot of the radii themselves as the central proton density is changed.
 
\begin{figure}[ht]
 \includegraphics[scale=0.7]{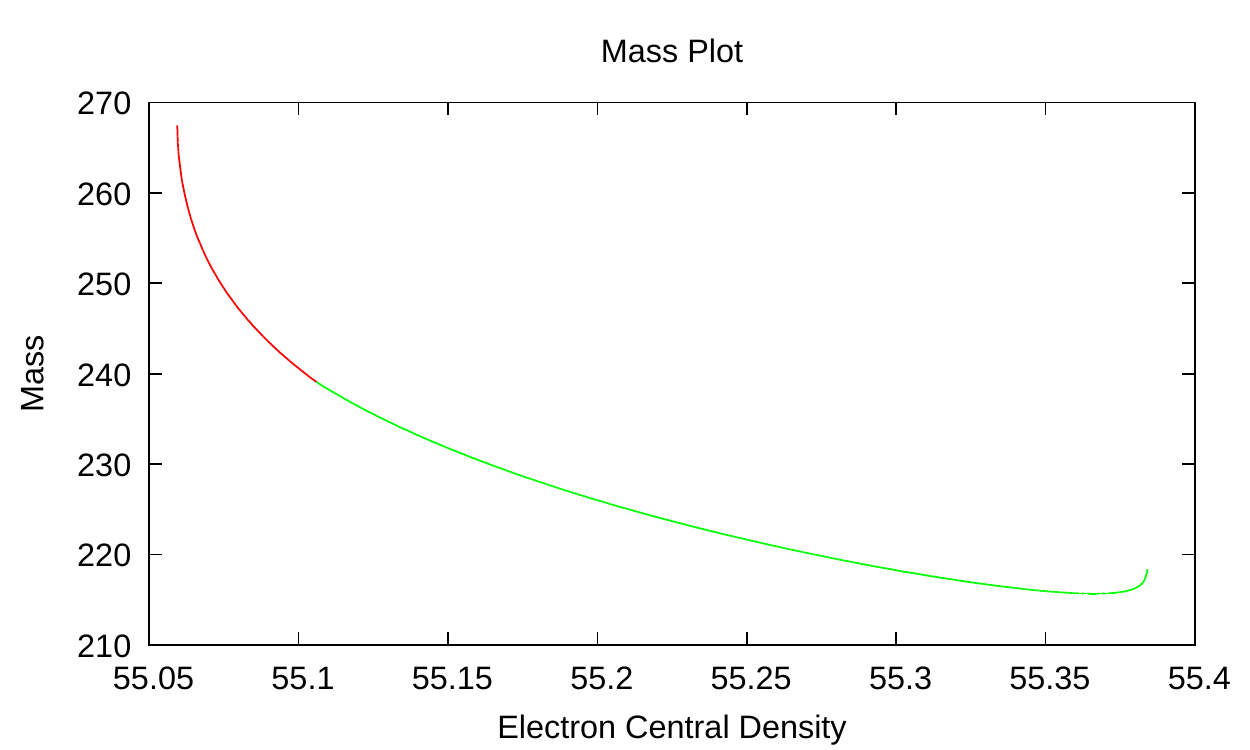}
\caption{Shown are the masses of the configurations as the central electron density is varied while the central proton density is held fixed at 100. The red colored part of the curve represents configurations with a positively charged atmosphere while the green part of the curve represents configurations with a negatively charged atmosphere. The science fiction values of $Gm_p^2/e^2=1/2$ and $m_e/m_p=1/10$ are used, so the values of mass are not meaningful.}
\label{massplot}
\end{figure}

\begin{figure}[H]
 \includegraphics[scale=0.7]{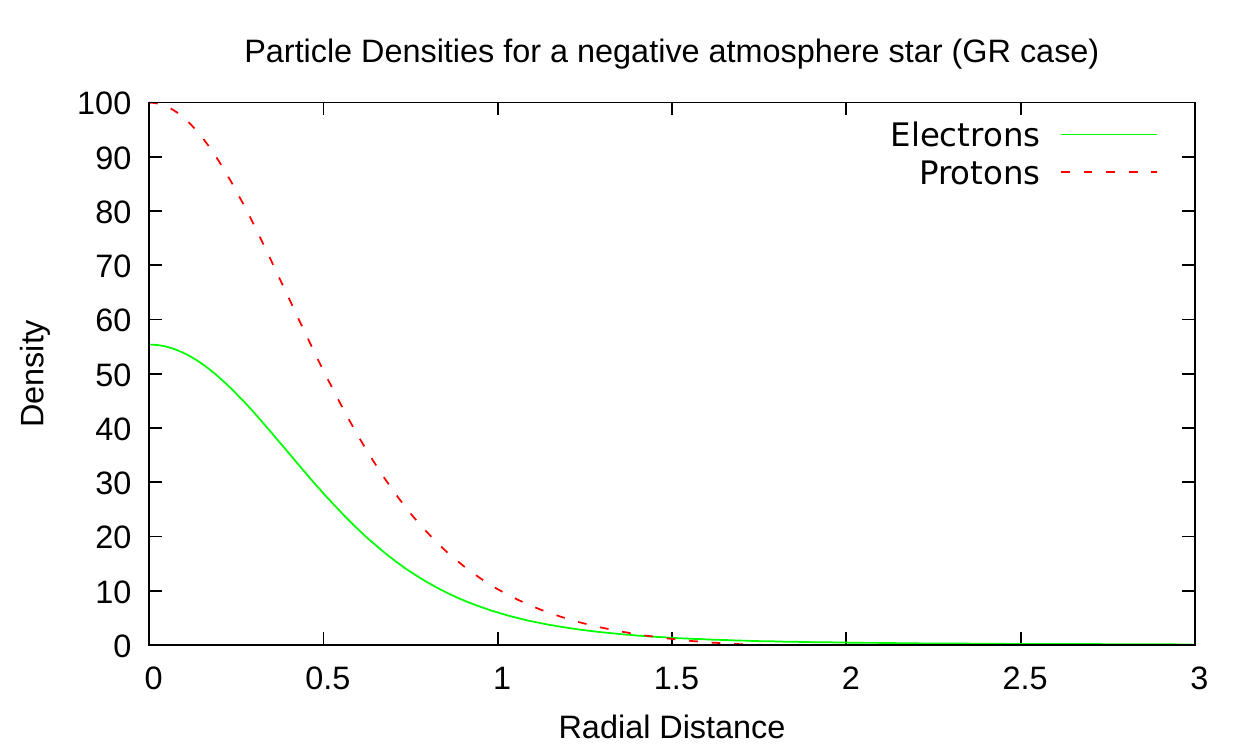}
  \includegraphics[scale=0.7]{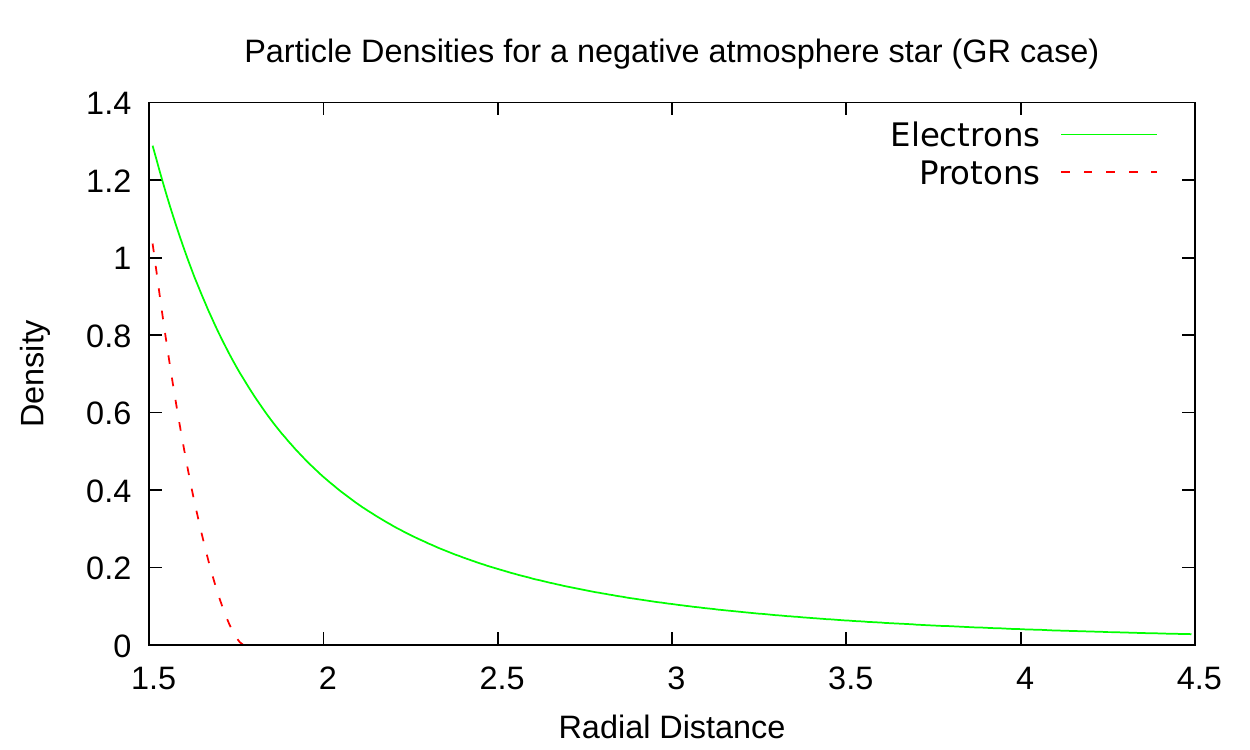}
\caption{Shown are the density functions for the protons and electrons of a configuration which is close to the lower bound on the charge and with science fiction values $Gm_p^2/e^2=1/2$ and $m_e/m_p=1/10$. The bottom graph is zoomed in on a part of the top graph.}
\label{GRneg}
\end{figure}

 Figure \ref{massplot} represents the same type of plot, but with the mass plotted instead of the radius. We say the star has a ``positively charged atmosphere'' when the support of the protons contains that of the electrons; otherwise, we say it has a ``negatively charged atmosphere''.
 
\begin{figure}[ht]
 \includegraphics[scale=0.7]{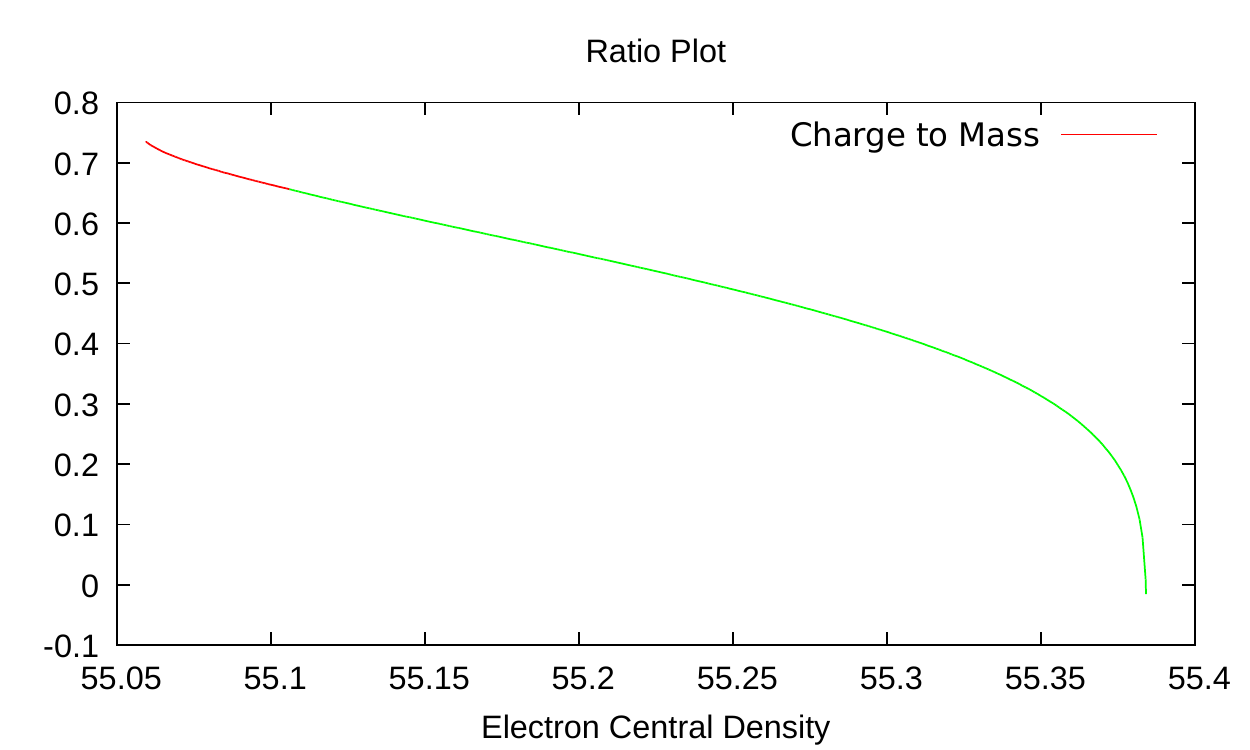}
\caption{Shown are the charge to mass ratios of the configurations as the central electron density is varied while the central proton density is held fixed at 100. The red colored part of the curve represents configurations with a positively charged atmosphere while the green part of the curve represents configurations with a negatively charged atmosphere. The science fiction values of $Gm_p^2/e^2=1/2$ and $m_e/m_p=1/10$ are used. It should be noted that, in these units, the minimum ratio is $\approx -0.0707$ while the maximum is $\approx 0.707$.}
\label{ratioplot}
\end{figure}

\begin{figure}[H]
 \includegraphics[scale=0.7]{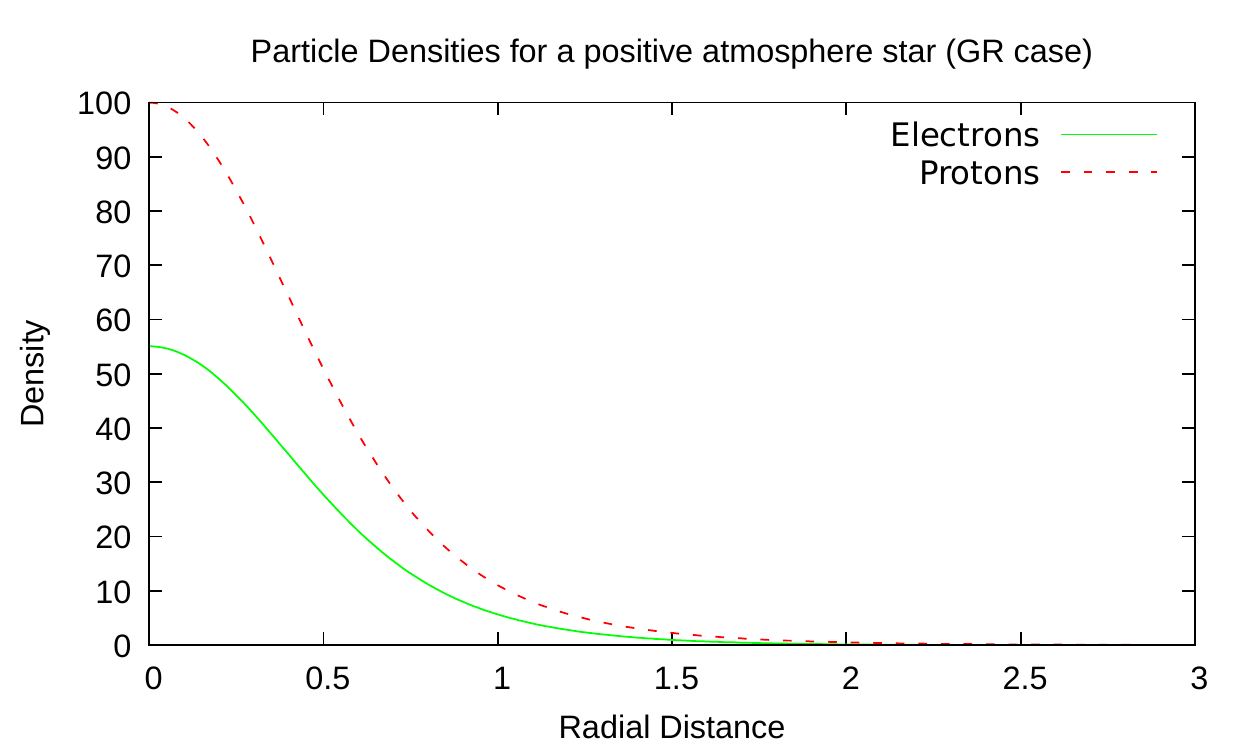}
  \includegraphics[scale=0.7]{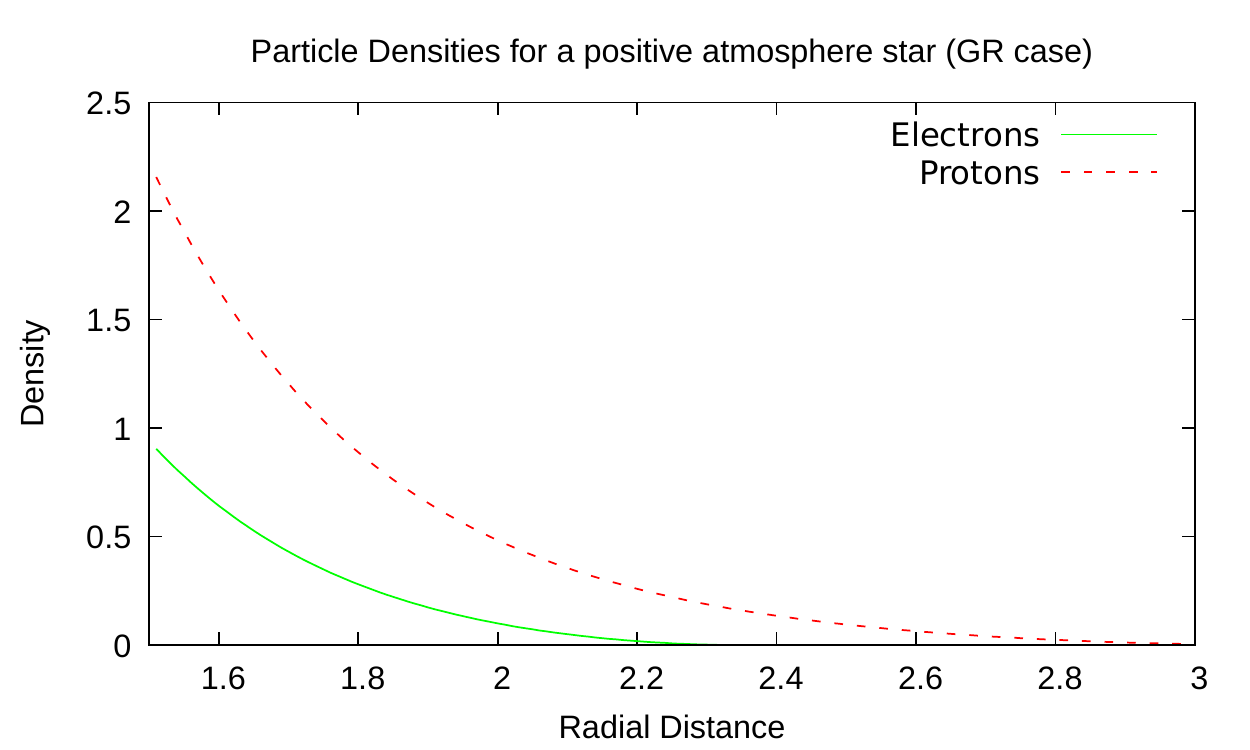}
\caption{Shown are the density functions for the protons and electrons of a configuration which is close to the upper bound on the charge and with science fiction values $Gm_p^2/e^2=1/2$ and $m_e/m_p=1/10$. The bottom graph is zoomed in on a part of the top graph.}
\label{GRpos}
\end{figure}

 Finally, figure \ref{ratioplot} plots the ratio of the total charge to the total mass. In section \ref{boundsSection}, it was derived that the bounds of this ratio are given by 
\begin{equation} \label{final_charge_bounds}
    -\frac{m_eG}{e}\leq\frac{Q}{M}\leq \frac{m_pG}{e} \ ,
\end{equation}
which in the science fiction values we are using is given approximately by  
\begin{equation}
    -0.0707\leq \frac{Q}{M}\leq 0.707 \ ,
\end{equation}
and we can see from the plot that we can get arbitrarily close to these values with numerical solutions.

\newpage

\begin{figure}[H]
 \includegraphics[scale=0.7]{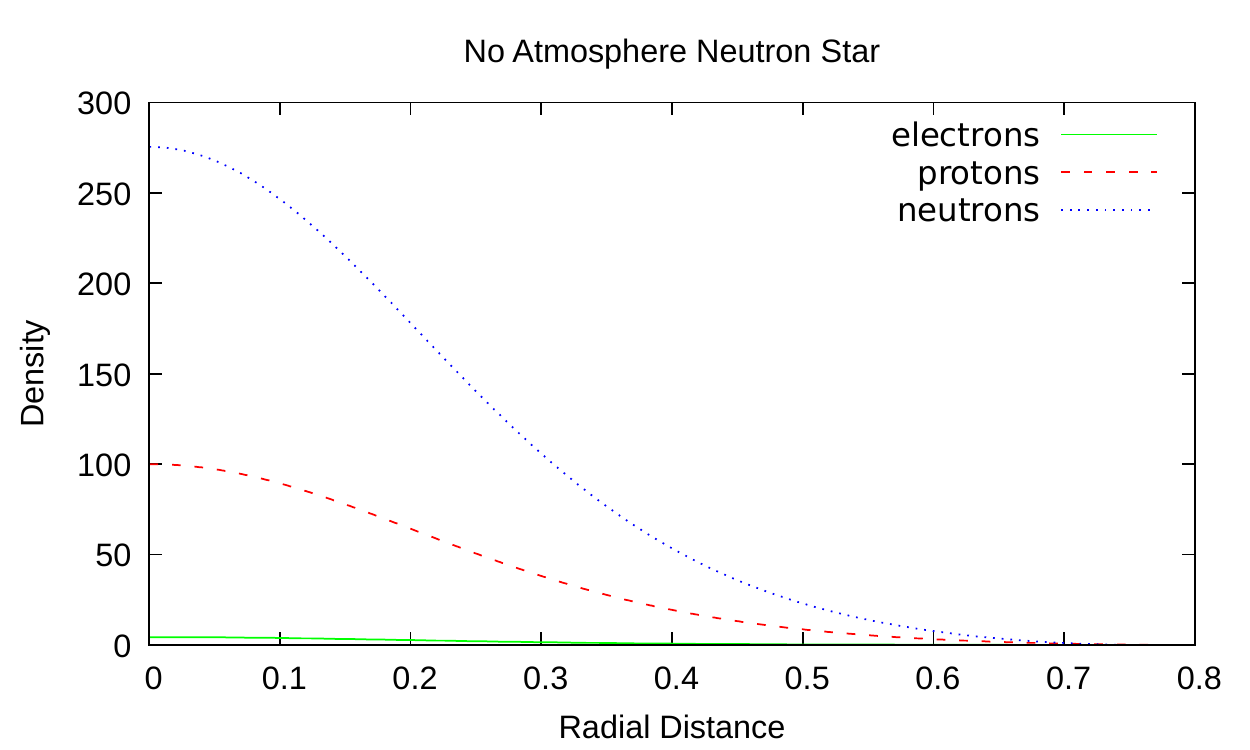}
  \includegraphics[scale=0.7]{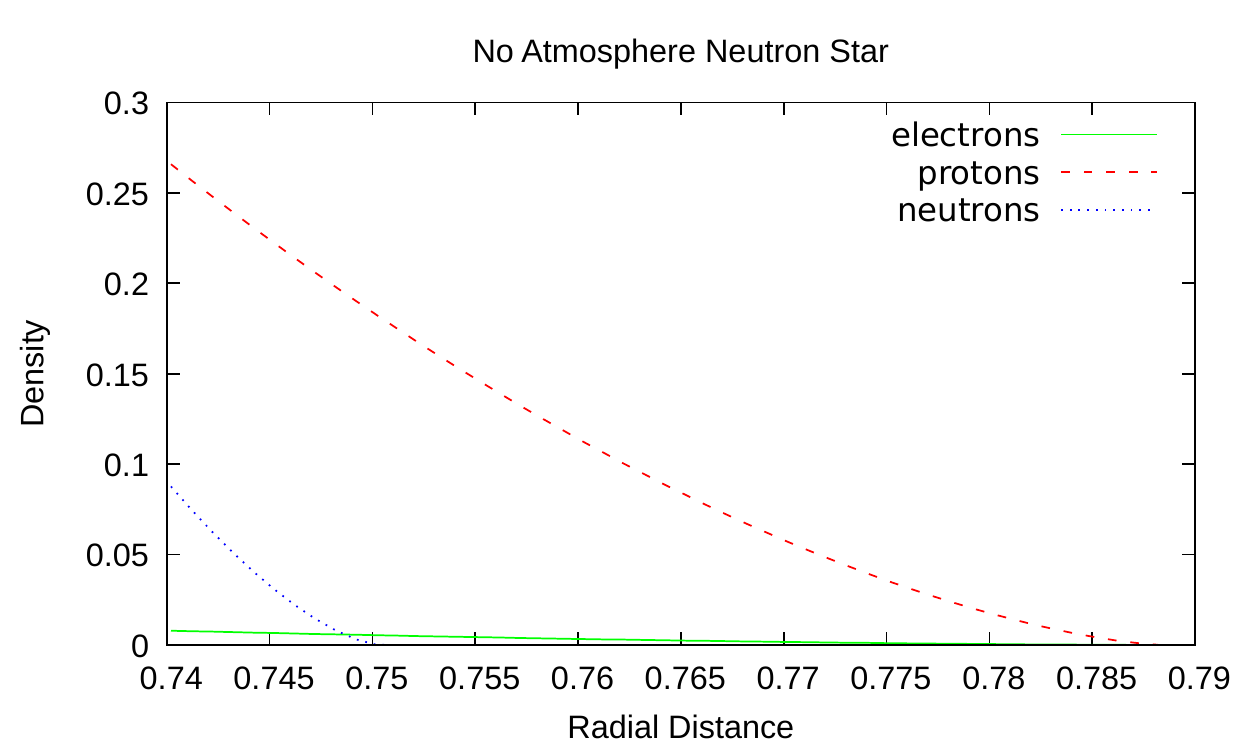}
\caption{Shown are the density functions for the protons, electrons, and neutrons of a neutron star configuration which has no atmosphere of protons or electrons and with science fiction values $Gm_p^2/e^2=1/2$ and $m_e/m_p=1/10$. The bottom graph is zoomed in on a part of the top graph.}
\label{NSnoatmos}
\end{figure}

\begin{figure}[ht]
 \includegraphics[scale=0.7]{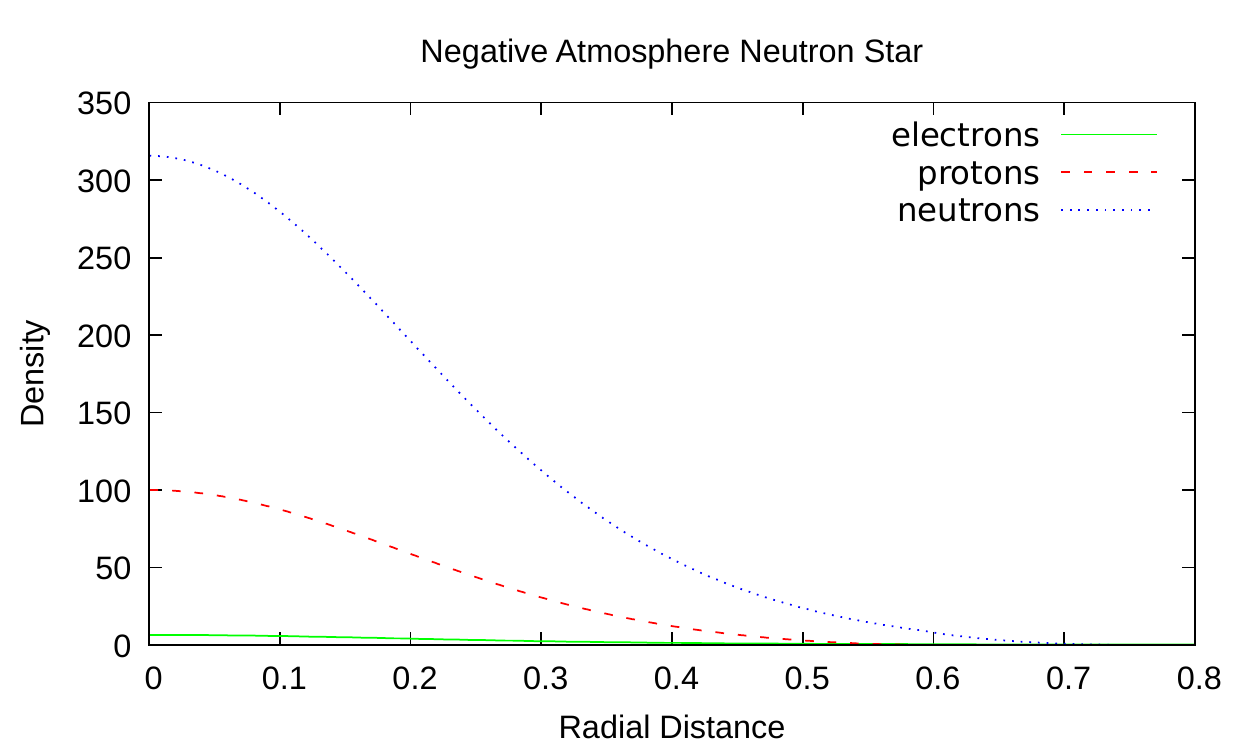}
  \includegraphics[scale=0.7]{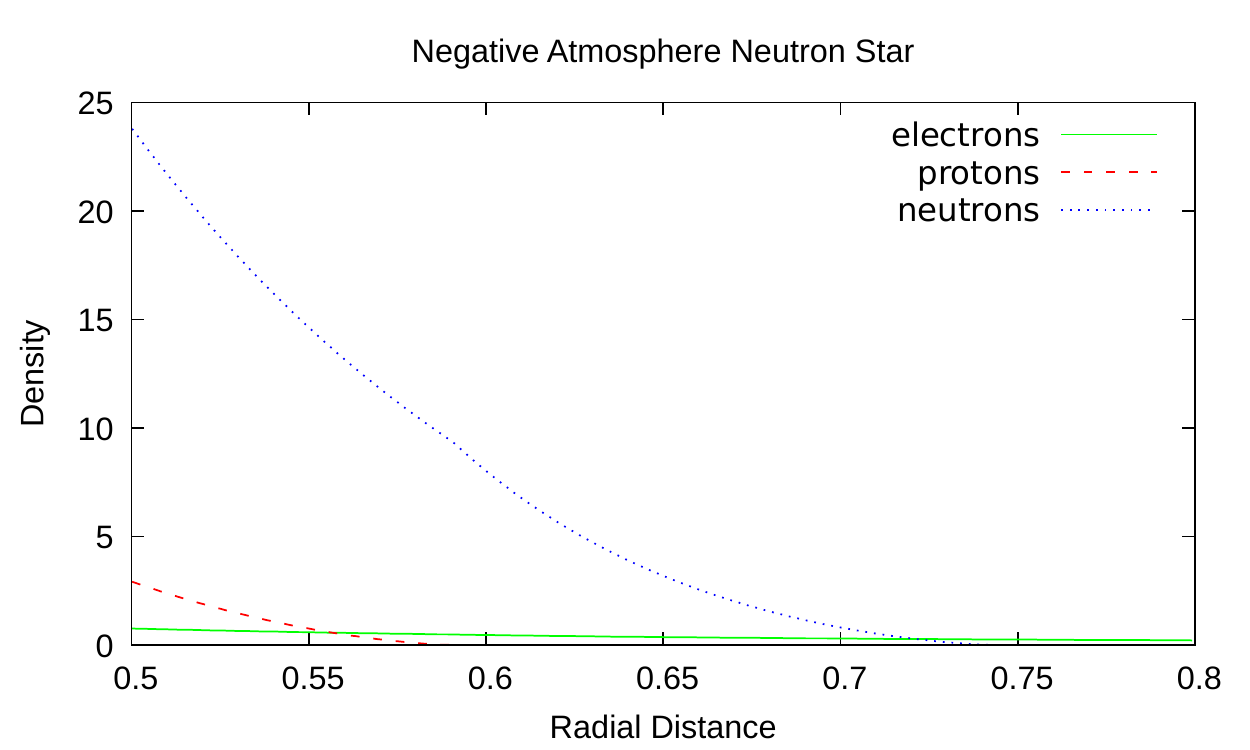}
\caption{Shown are the density functions for the protons, electrons, and neutron of a neutron star configuration which is close to the lower bound on the charge and with science fiction values $Gm_p^2/e^2=1/2$ and $m_e/m_p=1/10$. The bottom graph is zoomed in on a part of the top graph.}
\label{NSnegatmos}
\end{figure}
\begin{figure}[ht]
 \includegraphics[scale=0.7]{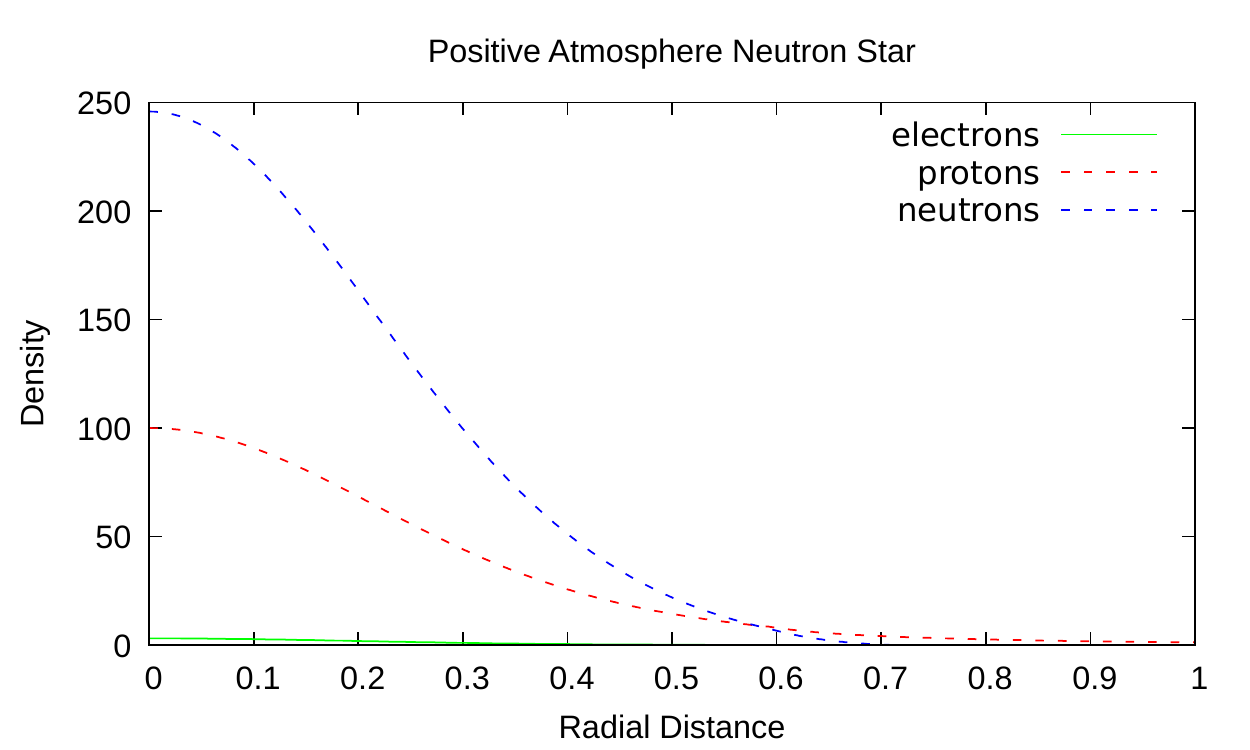}
  \includegraphics[scale=0.7]{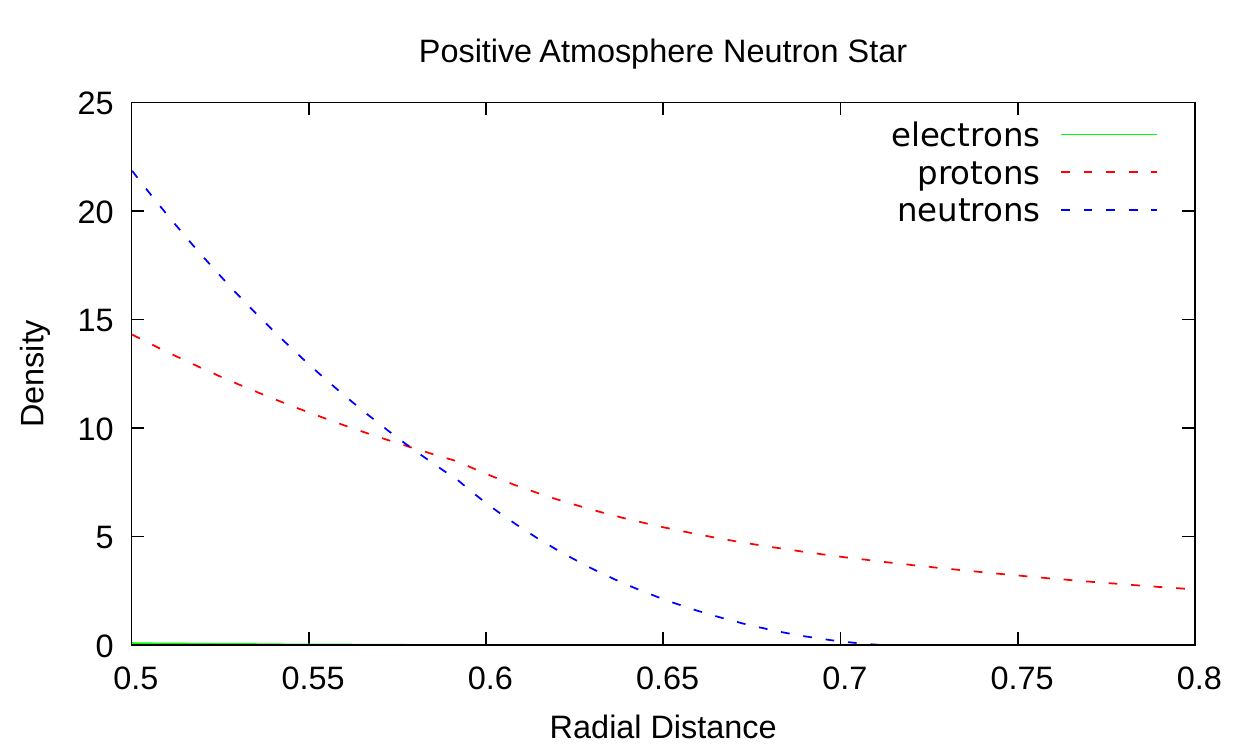}
\caption{Shown are the density functions for the protons, electrons, and neutrons of a neutron star configuration which is close to the upper bound on the charge and with science fiction values $Gm_p^2/e^2=1/2$ and $m_e/m_p=1/10$. The bottom graph is zoomed in on a part of the top graph.}
\label{NSposatmos}
\end{figure}

\section{Neutron Stars}
Everything we have done above applies as well when we consider neutron stars. The only difference in the neutron star model is that a third equation of the form (\ref{TOV2}) is added for the Fermi gas of neutrons. This third equation is implied by the equation for $\beta$-equilibrium 
\begin{equation}
    \mu_n=\mu_p+\mu_e \ .
\end{equation}
Figures \ref{NSnoatmos}, \ref{NSnegatmos}, and \ref{NSposatmos} illustrate, using the science-fiction values, some possible configurations given by this model. Immediately one notices that these science fiction values produce some rather extreme looking configurations: for a central proton density of 100, the largest the central electron density can be for a stable configuration is about 6.5. We have kept these same values for consistency, and also because we are interested in how the solutions change as we change the central densities, and, as in the white dwarf case, these science fiction values dramatically exaggerate the changes. Compare, for example, these plots to figure 2 in \cite{Ruffini2011a}.

Another thing one notices is that, as one approaches the extremely charged cases, the neutron density has a larger support than one of the electron or proton density. This would seem to indicate a limitation of the model.


\section{Conclusion}

We have applied the model developed by Olson and Bailyn in \cite{OB1}, \cite{OB2} and \cite{OB3} to write the equations of a general relativistic model of a star composed of fluids of electrons, protons and (optionally) neutrons. It generalizes the Newtonian model from \cite{KNY} and the special relativistic model from \cite{HK1} in the sense that all three make the same predictions in the Newtonian limit (large $c$), but we have also mentioned some differences between them in the post-Minkowski (small $G$) and weak field (flat metric) approximations. After pointing out an arithmetic error in the Olson-Bailyn papers which was inconsequential in their numeric calculations but relevant in ours, we showed how the equations of the model can be used to rigorously prove the bounds~\eqref{final_charge_bounds} on the total charge that the star can hold. The bounds proved were the same as in \cite{HK1} for the special relativistic model. We then plotted numerical solutions that do not assume local charge neutrality or approximate neutrality, as was the case in \cite{OB3}. The graphs obtained behave similarly to the special relativistic ones appearing in \cite{HK1}, and they also provide evidence that our charge bounds are close to begin sharp.

\hspace{5cm}
\section*{Appendix}
\subsection{Singularity of (\ref{OB3equation1})}

One could rewrite equation~\eqref{OB3equation1} in terms of the unknown
\begin{equation}
    s = \frac{h^2}{m_e^2c^2}\left(\frac{3n_e}{8\pi}\right)^{2/3}
\end{equation}
in the following form:
\begin{widetext}
\begin{align}
    &\frac{b+a^2}{K+a^2}+\frac{r}{K+a^2}\left(\frac{\partial b}{\partial r}+\frac{\partial b}{\partial s}s'+2a\left(\frac{\partial a}{\partial r}+\frac{\partial a}{\partial s}s'+\frac{\partial a}{\partial s'}s''\right)\right)\\
    &-r\frac{b+a^2}{(K+a^2)^2}\left(\frac{\partial K}{\partial r}+\frac{\partial K}{\partial s}s'+\frac{\partial K}{\partial s'}s''+2a\left(\frac{\partial a}{\partial r}+\frac{\partial a}{\partial s}s'+\frac{\partial a}{\partial s'}s''\right)\right)=\frac{8\pi Gr^2}{c^2}\rho_m+a^2\frac{K-b}{K+a^2}\nonumber
\end{align}
or, solving for $s''$,
\begin{align}\label{singularity}
    &\left(\frac{2ar}{K+a^2}-r\frac{b+a^2}{(K+a^2)^2}\left(\frac{\partial K}{\partial s'}+2a\frac{\partial a}{\partial s'}\right)\right)s''=-\frac{b+a^2}{K+a^2}-\frac{r}{K+a^2}\left(\frac{\partial b}{\partial r}+\frac{\partial b}{\partial s}s'+2a\left(\frac{\partial a}{\partial r}+\frac{\partial a}{\partial s}s'\right)\right)\\
    &+r\frac{b+a^2}{(K+a^2)^2}\left(\frac{\partial K}{\partial r}+\frac{\partial K}{\partial s}s'+2a\left(\frac{\partial a}{\partial r}+\frac{\partial a}{\partial s}s'\right)\right)+\frac{8\pi Gr^2}{c^2}\rho_m+a^2\frac{K-b}{K+a^2} \ . \nonumber
\end{align}
\end{widetext}
A simple heuristic way to see the singularity at $r=0$ is to consider the asymptotic behavior of these terms as $r\rightarrow0$. We assume $s(0)\neq0$, but that $s'(0)=0$, although we do not know the rate at which $s'$ goes to zero: by L'Hopital's rule we would need to know $s''$. Thus we have the following behaviors near zero: $a\sim rs'$, $b\sim r^2$, $K\sim 1$, $\frac{\partial a}{\partial s}\sim rs'$, $\frac{\partial a}{\partial r}\sim s'$, $\frac{\partial a}{\partial s'}\sim r$, $\frac{\partial b}{\partial s}\sim r^2$, $\frac{\partial b}{\partial r}\sim r$, $\frac{\partial K}{\partial r}\sim s'$, $\frac{\partial K}{\partial s}\sim rs'$, and $\frac{\partial K}{\partial s'}\sim r$.

One can check that the coefficient of $s''$ has leading order term behaving like $r^3s'$ or $r^4$: it is possible these terms could cancel, but then the coefficient would go to zero even faster. On the right side of (\ref{singularity}) however, the leading order terms are $-\frac{b}{K+a^2}$ and $-\frac{r}{K+a^2}\frac{\partial b}{\partial r}$ which do not cancel and behave like $r^2$ near zero. So there is no way to remove the singularity: to leading order we have either
\begin{equation}
    s''=\frac{c}{r^2}
\end{equation}
or 
\begin{equation}
    s''=\frac{c}{rs'}
\end{equation}
In the former case, $s$ has a singularity at $r=0$ from a logarithmic term. In the latter case, one can consider $T(r)=\frac{(s'(r))^2}{2}$ so that $T'(r)=s's''=\frac{c}{r}$, and $T(0)=0$. Fixing $r_0>0$, and $0<r<r_0$, we compute 
\begin{equation}
    T(r_0)-T(r)=\int_r^{r_0}T'(x)dx=c\ln(r_0/r) \ ,
\end{equation}
which implies that 
\begin{equation}
    T(r)=T(r_0)-c\ln(r_0/r) \ .
\end{equation}
So we must have $s'$ become unbounded around $r=0$.

\subsection{Computation of $\dfrac{\partial M}{\partial N_p}$}

Here we check inequality~\eqref{partialM_partialNp}. This calculation is quite similar to the computation of $\frac{\delta \Lambda}{\delta N_p}$ found in the appendix of \cite{OB1}. We remind the reader that the symbol $N_i$ stands for $N_i(R)$, the total number of particles of species $i$. We let $\delta$ denote variation with respect to the functions $N_i(r)$ defined by
\begin{equation} \label{app_Ni}
    N_i(r) = 4\pi\int_0^r n_i e^{\lambda/2}s^2ds \ ,
\end{equation}
assuming that $N_i(0)$ is held fixed. The label $i$ can mean $e$ or $p$. Start with 
\begin{equation} \label{app_delta_m}
    \delta m(r)=\frac{4\pi G}{c^2}\int_0^r\left(\delta \rho_m+\frac{2\mathcal{E}\delta\mathcal{E}}{8\pi c^2s^4}\right)s^2ds \ ,
\end{equation}
which comes from equation (3.4) in \cite{OB1}. We have 
\begin{equation} \label{app_delta_rho}
    \delta \rho_m=\frac{\partial \rho_m}{\partial n_p}\delta n_p+\frac{\partial \rho_m}{\partial n_e}\delta n_e \ .
\end{equation}
Differentiating~\eqref{app_Ni} and taking a variation, we find
\begin{equation}
    \delta n_i=\frac{1}{4\pi r^2}(e^{-\lambda/2}\delta N_i'+N_i'\delta e^{-\lambda/2}) \ ,
\end{equation}
while equation~\eqref{mass} gives
\begin{equation}
    \delta e^{-\lambda/2}=-e^{\lambda/2}\frac{\delta m(r)}{r} \ .
\end{equation}
\begin{widetext}
Plugging this into~\eqref{app_delta_rho}, we obtain
\begin{equation}
   \delta \rho_m=\frac{\partial \rho_m}{\partial n_p}\frac{1}{4\pi r^2}\left(e^{-\lambda/2}\delta N_p'-N_p'e^{\lambda/2}\frac{\delta m(r)}{r}\right)+\frac{\partial \rho_m}{\partial n_e} \frac{1}{4\pi r^2}\left(e^{-\lambda/2}\delta N_e'-N_e'e^{\lambda/2}\frac{\delta m(r)}{r}\right) \ .
\end{equation}
Then~\eqref{app_delta_m} becomes
\begin{align}
    \delta m(r)&=\frac{4\pi G}{c^2}\int_0^r\left[\frac{\partial \rho_m}{\partial n_p}\frac{1}{4\pi}\left(e^{-\lambda/2}\delta N_p'-N_p'e^{\lambda/2}\frac{\delta m(s)}{s}\right)+\frac{\partial \rho_m}{\partial n_e} \frac{1}{4\pi}\left(e^{-\lambda/2}\delta N_e'-N_e'e^{\lambda/2}\frac{\delta m(s)}{s}\right)\right] ds \\
    &+\frac{4\pi G}{c^2}\int_0^r\left[\frac{2\mathcal{E}}{8\pi c^2s^2}(q_p\delta N_p+q_e\delta N_e)\right]ds \ , \nonumber
\end{align}
where we have used~\eqref{calE_from_n} to write $\delta\mathcal{E}$ in terms of $\delta N_e$ and $\delta N_p$. Taking a derivative of both sides,
\begin{align}
    (\delta m)'+\frac{Ge^{\lambda/2}}{c^2r}\left(\frac{\partial \rho_m}{\partial n_p}N_p'+\frac{\partial \rho_m}{\partial n_e}N_e'\right)\delta m=\frac{G}{c^2}\left[e^{-\lambda/2}\left(\delta N'_p\frac{\partial \rho_m}{\partial n_p}+\delta N'_e\frac{\partial \rho_m}{\partial n_e}\right)+\frac{\mathcal{E}}{c^2r^2}(q_p\delta N_p+q_e\delta N_e)\right] \ .
\end{align}
As a solution we get
\begin{equation}
    \delta m(r)=e^{-D(r)}\int_0^r e^{D(s)}\frac{G}{c^2}\left[e^{-\lambda/2}\left(\delta N'_p\frac{\partial \rho_m}{\partial n_p}+\delta N'_e\frac{\partial \rho_m}{\partial n_e}\right)+\frac{\mathcal{E}}{c^2s^2}(q_p\delta N_p+q_e\delta N_e)\right] ds \ ,
\end{equation}
where we defined the integrating factor
\begin{equation} \label{int_factor}
    D(r)=\int_0^r\frac{Ge^{\lambda/2}}{c^2s}\left(\frac{\partial \rho_m}{\partial n_p}N_p'+\frac{\partial \rho_m}{\partial n_e}N_e'\right) ds \ .
\end{equation}
Since our goal is to find $\partial m(R)/\partial N_p$, assume only changes in the proton density from now on. We evaluate the above at $r=R$ and use integration by parts:
\begin{equation}
   \delta m(R) = e^{-D(R)}\frac{G}{c^2}\int_0^R \left[e^{D(s)}\frac{\mathcal{E}q_p}{c^2s^2}\delta N_p-\left(e^{D(s)}e^{-\lambda/2}\frac{\partial \rho_m}{\partial n_p}\right)'\delta N_p\right] ds+\frac{G}{c^2}e^{-\lambda(R)/2}\frac{\partial \rho_m}{\partial n_p}(R)\delta N_p(R) \ .
\end{equation}
The integrand of the first term on the right hand side evaluates to zero with the help of the definition of $D$ in~\eqref{int_factor} and of the species TOV equation~\eqref{TOV2} for $(\frac{\partial\rho_m}{\partial n_p})' = \frac{\mu_p'}{c^2}$. Therefore
\begin{equation}
    \delta m(R) = \frac{G}{c^2}e^{-\lambda(R)/2}\frac{\partial \rho_m}{\partial n_p}(R)\delta N_p \ ,
\end{equation}
which implies that
\begin{equation}
    \frac{\partial m(R)}{\partial N_p} = \frac{G}{c^2}e^{-\lambda(R)/2}\frac{\partial \rho_m}{\partial n_p}(R) \ .
\end{equation}

Taking into account $M = \frac{c^2}{G}m$ and $m_pc^2 = \mu_p(R) = c^2\frac{\partial\rho_m}{\partial n_p}(R)$, we reach the conclusion that
\begin{equation}
    \frac{\partial M}{\partial N_p} = m_p e^{-\lambda(R)/2} > 0 \ .
\end{equation}

\end{widetext}

\section*{Acknowledgment}
 We thank Michael Kiessling, Shadi Tahvildar-Zadeh, and Eric Ling for helpful discussions.

\bibliographystyle{apsrev}

ph325@math.rutgers.edu 

edeamori@math.uni-koeln.de

\end{document}